\documentclass{jss}

\usepackage[utf8]{inputenc}

\providecommand{\tightlist}{%
  \setlength{\itemsep}{0pt}\setlength{\parskip}{0pt}}

\author{
Emma C. Martin\\University of Leicester \And Alessandro Gasparini\\Karolinska Institutet \And Michael J. Crowther\\University of Leicester \&\\
Karolinska Institutet
}
\title{\pkg{merlin}: An R package for Mixed Effects Regression for Linear,
Nonlinear and User-defined models}

\Plainauthor{Emma C. Martin, Alessandro Gasparini, Michael J. Crowther}
\Plaintitle{merlin: An R package for Mixed Effects Regression for Linear, Nonlinear
and User-defined models}
\Shorttitle{merlin: Mixed Effects Regression}

\Abstract{
The \proglang{R} package \pkg{merlin} performs flexible joint modelling
of hierarchical multi-outcome data. Increasingly, multiple longitudinal
biomarker measurements, possibly censored time-to-event outcomes and
baseline characteristics are available. However, there is limited
software that allows all of this information to be incorporated into one
model. In this paper, we present \pkg{merlin} which allows for the
estimation of models with unlimited numbers of continuous, binary, count
and time-to-event outcomes, with unlimited levels of nested random
effects. A wide variety of link functions, including the expected value,
the gradient and shared random effects, are available in order to link
the different outcomes in a biologically plausible way. The accompanying
\code{predict.merlin} function allows for individual and population
level predictions to be made from even the most complex models. There is
the option to specify user-defined families, making \pkg{merlin} ideal
for methodological research. The flexibility of \pkg{merlin} is
illustrated using an example in patients followed up after heart valve
replacement, beginning with a linear model, and finishing with a joint
multiple longitudinal and competing risks survival model.
}

\Plainkeywords{joint modelling, multi-outcome, R, merlin}

\Address{
    Emma C. Martin\\
  University of Leicester\\
  Department of Health Sciences, University Road, Leicester, LE1 7RH, UK\\
  E-mail: \email{emma.martin@le.ac.uk}\\

      Alessandro Gasparini\\
  Karolinska Institutet\\
  Department of Medical Epidemiology and Biostatistics, Stockholm, Sweden\\
  E-mail: \email{alessandro.gasparini@ki.se}\\

      Michael J. Crowther\\
  University of Leicester \&\\
  Karolinska Institutet\\
  Department of Health Sciences, University Road, Leicester, LE1 7RH, UK
  \& Department of Medical Epidemiology and Biostatistics, Stockholm,
  Sweden\\
  E-mail: \email{michael.crowther@le.ac.uk}\\

  }

\usepackage{amsmath} \usepackage{thumbpdf} \usepackage{lmodern}

\begin{document}

\hypertarget{introduction}{%
\section{Introduction}\label{introduction}}

Software packages to fit joint and multi-state models are continuously
being developed and updated to increase flexibility. However this
flexibility is often limited in terms of outcome types, levels of nested
random-effects, or the forms of linking functions between outcomes. We
have developed \pkg{merlin} in order to address this lack of
flexibility, allowing for a wide range of models to be estimated. With
\pkg{merlin} it is possible to include any number of outcomes from a
wide range of families, including Gaussian, Bernoulli, Poisson, a number
of survival models including flexible parametric models, amongst others.
It is also possible to custom supply user defined families to allow for
even greater flexibility and method development. This allows
\pkg{merlin} to fit everything from a simple Weibull model to a
multivariate joint model. Joint models can be defined using commonly
chosen association structures \citep{Gould2015a}, for example, shared
random effects, the current value, gradient or area under the curve, and
to provide even more customisation - user defined link functions. This
\proglang{R} package is based on the recently released \pkg{merlin}
package in Stata \citep[\citet{merlinStata}]{Crowther2017}.

Previous software released in \proglang{R} has some of the individual
capabilities of \pkg{merlin}. Package \pkg{JM} \citep{JM} fits a single
normal longitudinal response jointly with a single survival outcome or
competing risk outcomes, assuming a current value or current gradient
link. There is also an extension \pkg{JMbayes} \citep{JMbayes} which
fits similar models in a Bayesian framework. \pkg{joineR} \citep{joineR}
allows for the joint modelling of a single longitudinal response and a
single time-to-event outcome or competing risk outcome. The extension
\pkg{joineRML} \citep{joineRML} additionally allows for multivariate
longitudinal data. The \pkg{frailtypack} \citep{frailtypack} package
fits shared, joint and nested frailty models, with one longitudinal
response and multiple recurrent and terminal events.

New package \pkg{merlin} offers additional flexibility in how the joint
model is specified. Multiple longitudinal responses can be specified and
there is a wider range of models available to better describe the data,
including splines and fractional polynomials. There is also a wider
variety of survival models available compared to \pkg{joineR} which only
allows Cox models, and \pkg{JM} and \pkg{frailtypack} which allow Cox,
Weibull and limited spline based survival models. In addition to these
models, \pkg{merlin} allows for exponential survival models, and a wider
range of flexible spline based models such as Royston-Parmar models
\citep{Royston2001}. While it is possible to fit models with multiple
shared random-effects, there are additional link functions available to
describe the relationship between the longitudinal and time-to-event
outcomes, including current expected value, or other functions of the
longitudinal response, including derivatives and integrals. A number of
packages exist which allow for multiple hierarchical levels of
random-effects for either longitudinal responses (\pkg{lme4}
\citep{lme4} or \pkg{nlme} \citep{nlme}) or time-to-event outcomes
(\pkg{coxme} \citep{coxme}). Each of the joint modelling packages
described above only allow for one level of clustering, with the
exception of \pkg{frailtypack} which allows for two, whereas
\pkg{merlin} can incorporate any number of nested levels, which is
particularly useful for big data such as electronic health records,
which is often hierarchical.

Further flexibility is provided in \pkg{merlin} with the option of
user-defined functions. This allows users to define their own likelihood
functions, \pkg{merlin} is then used as a wrapper function to carry out
the optimisation, similar to \pkg{BAMLSS} \citep{BAMLSS} which uses a
modular ``Lego brick'' approach in a Bayesian framework. Allowing users
to extend \pkg{merlin} via user-defined functions makes it a useful tool
for the development of new methods.

In this paper we introduce the modular syntax employed by \pkg{merlin}
which enables its flexibility. In order to illustrate this flexibility
we will develop an example model using data from an observational study
of patients following aortic valve replacement surgery \citep{Lim2008}.
In Section 2 we explain the syntax to specify the model structure and
use the predict function. In Section 3 we work through an illustrative
example in patients following heart valve replacement. Finally in
Section 4 we discuss the advantages of using \pkg{merlin} and plans for
future extensions.

\hypertarget{specifying-model-structure}{%
\section{Specifying model structure}\label{specifying-model-structure}}

\hypertarget{syntax}{%
\subsection{Syntax}\label{syntax}}

The syntax for \pkg{merlin} is modular in nature. The family is
specified for each outcome, the linear predictor for each outcome can
then be built from components such as an intercept, covariates and
random effects.

\begin{verbatim}
merlin(model = list(model1, model2, ...),
       family = c("family1", "family2", ...),
       levels = "level1",
       data = data))
\end{verbatim}

Where the syntax for each model is

\code{
model1 <- depvar ~ component1 + component2 + ..., model_options
}

Each component can be made up of a number of elements such as
covariates, random effects, functions of time and expected values of
other outcomes. Interactions between elements can be specified using
\code{:} between different elements. By default a coefficient will be
estimated for each component, the coefficient can be constrained to 1
using \code{*1}.

\code{
component1 <- element1 [:element2] [:element3] [...] [*1]
}

A number of model families are currently available, including

\begin{itemize}
\tightlist
\item
  \code{gaussian} - Gaussian distribution
\item
  \code{bernoulli} - Bernoulli distribution
\item
  \code{poisson} - Poisson distribution
\item
  \code{beta} - beta distribution
\item
  \code{negbinomial} - Negative binomial distribution
\end{itemize}

As well as a number of survival models

\begin{itemize}
\tightlist
\item
  \code{exponential} - exponential survival distribution
\item
  \code{weibull} - Weibull distribution
\item
  \code{gompertz} - Gompertz distribution
\item
  \code{rp} - Royston-Parmar survival model, (complex predictor on the
  log cumulative hazard scale)
\item
  \code{loghazard} - general log hazard model (complex predictor on the
  log hazard scale)
\end{itemize}

With two user-defined options

\begin{itemize}
\tightlist
\item
  \code{user} - which fits a user-defined model which can be written
  using \pkg{merlin}'s utility functions. The name of the user-defined
  function needs to be passed through using \code{userf} option
\item
  \code{null} - which is a convenience tool for defining additional
  complex predictors, that do not contribute to the log likelihood
\end{itemize}

\hypertarget{element-types}{%
\subsection{Element types}\label{element-types}}

Each element can take a number of different forms

\begin{itemize}
\tightlist
\item
  \code{varname} - the simplest form is a varname, which refers to a
  variable in the data set provided.
\item
  \code{rcs} - a restricted cubic spline function,

  \begin{itemize}
  \tightlist
  \item
    \code{knots()} - allows the user to specify the location of the
    knots in the form of a vector.
  \item
    \code{df()} - alternatively the number of degrees of freedom can be
    specified, in which case the boundary knots are assumed to be at the
    minimum and maximum of \code{varname} with the internal knots placed
    at evenly spaced centiles.
  \item
    \code{orthog} - this option uses Gram-Schmidt orthogonalisation of
    the splines, specifying this can improve model convergence.
  \end{itemize}
\item
  \code{time functions} - such as powers of time and log time. In order
  to use time functions \code{timevar} must be specified as extra
  numerical integration may be required.
\item
  \code{M\#[cluster level]} - a random-effect at the cluster level, all
  random-effects must be named \code{M} followed by a number to enable
  the sharing of random effects between models
\item
  \code{fp()} - specifies a fractional polynomial function, with order 1
  or 2.

  \begin{itemize}
  \tightlist
  \item
    \code{powers()} - the powers of the the fractional polynomial
    function must be specified (up to second degree).
  \end{itemize}
\item
  \code{bhazard(varname)} - invokes a relative survival (excess hazard)
  model. \code{varname} specifies the expected hazard rate at the event
  time.
\item
  \code{exposure(varname)} - include log(\code{varname}) in the linear
  predictor, with a coefficient of 1. For use with
  \code{family = "poisson"}.
\end{itemize}

Functions of longitudinal submodels can be included as covariates in
other submodels using the following options

\begin{itemize}
\tightlist
\item
  \code{EV[depvar]} - the expected value of the response of a submodel
\item
  \code{dEV[depvar]} - the first derivative with respect to time of the
  expected value of the response of a submodel
\item
  \code{d2EV[depvar]} - the second derivative with respect to time of
  the expected value of the response of a submodel
\item
  \code{iEV[depvar]} - the integral with respect to time of the expected
  value of the response of a submodel
\item
  \code{XB[depvar]} - the expected value of the complex predictor of a
  submodel
\item
  \code{dXB[depvar]} - the first derivative with respect to time of the
  expected value of the complex predictor of a submodel
\item
  \code{d2XB[depvar]} - the second derivative with respect to time of
  the expected value of the complex predictor of a submodel
\item
  \code{iXB[depvar]} - the integral with respect to time of the expected
  value of the complex predictor of a submodel
\end{itemize}

\hypertarget{integration-methods}{%
\subsection{Integration methods}\label{integration-methods}}

There are a number of methods available for numerically integrating out
the random-effects in order to calculate the likelihood for a
mixed-effects model. The options for \code{intmethod} are:

\begin{itemize}
\tightlist
\item
  \code{ghermite} - for non-adaptive Gauss-Hermite quadrature;
\item
  \code{halton} - for Monte Carlo integration using Halton sequences;
\item
  \code{sobol} - for Monte Carlo integration using Sobol sequences;
\item
  \code{mc} - for standard Monte Carlo integration using normal draws.
\end{itemize}

The default is \code{ghermite}. Level-specific integration techniques
can be specified. Gauss-Hermite quadrature is widely considered the
optimal numerical integration technique, however it doesn't scale well
for large numbers of random-effects. Therefore in a three level model
example, we may use Gauss-Hermite quadrature at the highest level and
the more efficient Monte-Carlo integration with Halton sequences at
level 2, using \code{intmethod = c("ghermite", "halton")}.

\hypertarget{post-estimation}{%
\subsection{Post estimation}\label{post-estimation}}

A range of post estimation tools are available with \pkg{merlin} using
the prediction function using the following syntax.

\code{
predict(modelname, statistic, type, options)
}

The currently available statistics options are

\begin{itemize}
\tightlist
\item
  \code{eta} - the expected value of the complex predictor
\item
  \code{mu} - the expected value of the response variable
\item
  \code{hazard} - the hazard function
\item
  \code{chazard} - the cumulative hazard function
\item
  \code{logchazard} - the log cumulative hazard function
\item
  \code{survival} - the survival function
\item
  \code{cif} - the cumulative incidence function
\item
  \code{rmst} - calculates the restricted mean survival time, which is
  the integral of the survival function within the interval (0,t{]},
  where t is the time at which predictions are made. If multiple
  survival models have been specified in your \pkg{merlin} model, then
  it will assume all of them are cause-specific competing risks models,
  and include them in the calculation. If this is not the case, you can
  override which models are included by using the \code{causes} option.
  \code{rmst = t - totaltimelost}.
\item
  \code{cifdifference} calculates the difference in \code{cif}
  predictions between values of a covariate specified using the
  \code{contrast} option.
\item
  \code{hdifference} calculates the difference in \code{hazard}
  predictions between values of a covariate specified using the
  \code{contrast} option.
\item
  \code{rmstdifference} calculates the difference in \code{rmst}
  predictions between values of a covariate specified using the
  \code{contrast} option.
\item
  \code{mudifference} calculates the difference in \code{mu} predictions
  between values of a covariate specified using the \code{contrast}
  option.
\item
  \code{etadifference} calculates the difference in \code{eta}
  predictions between values of a covariate specified using the
  \code{contrast} option.
\item
  \code{timelost} - calculates the time lost due to a particular event
  occurring, within the interval (0,t{]}. In a single event survival
  model, this is the integral of the cif between (0,t{]}. If multiple
  survival models are specified in the \pkg{merlin} model then by
  default all are assumed to be cause-specific event time models
  contributing to the calculation. This can be overridden using the
  \code{causes} option.
\item
  \code{totaltimelost} - total time lost due to all competing events,
  within (0,t{]}. If multiple survival models are specified in the
  \pkg{merlin} model then by default all are assumed to be
  cause-specific event time models contributing to the calculation. This
  can be overridden using the \code{causes} option. \code{totaltimelost}
  is the sum of the \code{timelost} due to all causes.
\end{itemize}

Prediction options include

\begin{itemize}
\tightlist
\item
  \code{type} - specifies whether the predictions include fixed-effects
  only (\code{fixedonly}), or the marginal prediction is calculated
  marginally with respect to the latent variables. The \code{stat} is
  calculated by integrating the prediction function with respect to all
  the latent variables over their entire support.
\item
  \code{predmodel} - specifies which model to predict from, default
  \code{predmodel=1}.
\item
  \code{causes} - for use when calculating predictions from a competing
  risks model. By default, \code{cif}, \code{rmst}, \code{timelost} and
  \code{totaltimelost} assume that all survival models included in the
  \pkg{merlin} model are cause-specific hazard models contributing to
  the calculation. If this is not the case, then you can specify which
  models (indexed using the order they appear in your \pkg{merlin} model
  by using the \code{causes} option, e.g.~\code{causes=c(1, 2)}).
\item
  \code{at} - specifies covariate values for prediction. Fixed values of
  covariates should be specified in a list
  e.g.~\code{at = c("trt" = 1, "age" = 50)}.
\item
  \code{contrast} - specifies the values of a covariate to be used when
  comparing statistics, such as when using the \code{cifdifference}
  option to compare cumulative incidence functions,
  e.g.~\code{contrast = c("trt" = 0, "trt" = 1)}.
\end{itemize}

\hypertarget{examples}{%
\section{Examples}\label{examples}}

A consequence of the flexibility of \pkg{merlin} is the syntax is
arguably complex to allow for the generalisation. In order to illustrate
the potential uses of \pkg{merlin} we fit a number of increasingly
advanced models to data from an observational study which investigated
the effects of aortic valve replacement with a stentless or a homograft
valve \citep{Lim2008}. The study followed 300 patients who underwent
aortic valve replacement between 1991 and 2001, all patients with at
least one year of follow-up were included. A number of baseline
measurements were available such as age, sex, preoperative body surface
area and size of valve. The dataset also includes longitudinal measures
of valve gradient, standardised left ventricular mass index and ejection
fraction from an average of four follow-up appointments per patient. We
will use the copy of the data set available from \pkg{R} package
\pkg{joineRML} \citep{joineRML} to illustrate.

\begin{CodeChunk}

\begin{CodeInput}
R> data(heart.valve, package = "joineRML")
\end{CodeInput}
\end{CodeChunk}

As we are interested in fitting joint survival and longitudinal models,
there must be at least one longitudinal biomarker measurement for each
individual. We will primarily focus on valve gradient as our
longitudinal outcome, therefore it is necessary to exclude any
individual who doesn't have at least one valve gradient observation.

\begin{CodeChunk}

\begin{CodeInput}
R> heart.valve <- heart.valve[!is.na(heart.valve$grad), ]
\end{CodeInput}
\end{CodeChunk}

The data should be set out in wide format for submodel, with each
outcome specified in a separate column. For survival data event time and
status will appear in different columns, and should only be specified
once per individual. However within a submodel long format is used, with
each repeated measurement of a biomarker on a new row with a separate
column specifying the timing of that observation. The current set up of
the \code{heart.valve} data (shown below) needs some editing to allow
modelling fitting with \pkg{merlin}, however once the data has been put
into the correct format, all models can be fitted without any further
editing being required.

\begin{CodeChunk}

\begin{CodeInput}
R> print(
R+   heart.valve[heart.valve$num %in% c(1:2, 13),
R+               c(1:3, 25, 5:6, 4, 8, 10)],
R+   row.names = FALSE
R+ )
\end{CodeInput}

\begin{CodeOutput}
 num sex      age              hs    fuyrs status      time log.grad log.lvmi
   1   0 75.06027 Stentless valve 4.956164      0 0.0109589 2.302585 4.778955
   1   0 75.06027 Stentless valve 4.956164      0 3.6794520 2.302585 4.778955
   1   0 75.06027 Stentless valve 4.956164      0 4.6958900 2.302585 4.924569
   2   0 45.79452       Homograft 9.663014      0 6.3643840 2.639057 4.744323
   2   0 45.79452       Homograft 9.663014      0 7.3041100 2.197225 4.698661
   2   0 45.79452       Homograft 9.663014      0 8.3013700 2.484907 5.058790
  13   1 69.94247       Homograft 5.186301      1 0.1369863 2.708050 5.305541
  13   1 69.94247       Homograft 5.186301      1 1.0575340 2.833213 5.283356
  13   1 69.94247       Homograft 5.186301      1 2.0547950 2.995732 4.794550
  13   1 69.94247       Homograft 5.186301      1 3.9726030 3.401197 4.993421
\end{CodeOutput}
\end{CodeChunk}

The event time (\code{fuyrs}) for each individual should only appear
once in the data set, unless there are multiple events per individual,
these should appear on separate lines. The \code{status} is the event
indicator variable, coded 0 for censored (lost to follow-up) and 1 for
died.

\begin{CodeChunk}

\begin{CodeInput}
R> heart.valve$id <- heart.valve$num
R> heart.valve$stime <- heart.valve$fuyrs
R> heart.valve$stime[duplicated(heart.valve$id)] <- NA
R> heart.valve$died <- heart.valve$status
R> heart.valve$died[duplicated(heart.valve$id)] <- NA
\end{CodeInput}
\end{CodeChunk}

Binary variables, such as type of heart valve used, need to be converted
to be numeric.

\begin{CodeChunk}

\begin{CodeInput}
R> heart.valve$type <- as.numeric(heart.valve$hs) - 1
\end{CodeInput}
\end{CodeChunk}

A section of the correctly formatted data is shown below, Individual 1
has three longitudinal measurements for log valve gradient
(\code{log.grad}) and log left ventricular mass index (\code{log.lvmi}),
the timings of these measurements are given in the \code{time} column.
In this case the different biomarkers were measured at the same time
points, but this is not necessary, missing biomarker measurements should
be recorded as \code{NA}. The survival information has been recorded on
the first line for each individual. Baseline covariates such as sex
should be specified on every line for that individual. All the models
below are fitted to this data set.

\begin{CodeChunk}

\begin{CodeInput}
R> print(
R+   heart.valve[heart.valve$id %in% c(1:2, 13),
R+               c(26, 2:3, 29, 27:28, 8, 10, 4)],
R+   row.names = FALSE
R+ )
\end{CodeInput}

\begin{CodeOutput}
 id sex      age type    stime died log.grad log.lvmi      time
  1   0 75.06027    1 4.956164    0 2.302585 4.778955 0.0109589
  1   0 75.06027    1       NA   NA 2.302585 4.778955 3.6794520
  1   0 75.06027    1       NA   NA 2.302585 4.924569 4.6958900
  2   0 45.79452    0 9.663014    0 2.639057 4.744323 6.3643840
  2   0 45.79452    0       NA   NA 2.197225 4.698661 7.3041100
  2   0 45.79452    0       NA   NA 2.484907 5.058790 8.3013700
 13   1 69.94247    0 5.186301    1 2.708050 5.305541 0.1369863
 13   1 69.94247    0       NA   NA 2.833213 5.283356 1.0575340
 13   1 69.94247    0       NA   NA 2.995732 4.794550 2.0547950
 13   1 69.94247    0       NA   NA 3.401197 4.993421 3.9726030
\end{CodeOutput}
\end{CodeChunk}

\hypertarget{linear-regression}{%
\subsection{Linear regression}\label{linear-regression}}

To begin with we will fit a simple linear regression of log of the valve
gradient (\code{log.grad}) against \code{time}, with \code{age} and
\code{sex} as covariates.

\begin{CodeChunk}

\begin{CodeInput}
R> library(merlin)
R> m1 <- merlin(
R+   model = log.grad ~ sex + age + time,
R+   family = "gaussian",
R+   data = heart.valve
R+ )
R> summary(m1)
\end{CodeInput}

\begin{CodeOutput}
Mixed effects regression model
Log likelihood = -651.4753

                Estimate Std. Error       z Pr(>|z|) [95% Conf. Interval]
sex             0.140489   0.059778   2.350   0.0188   0.023327  0.257651
age            -0.002212   0.002297  -0.963   0.3356  -0.006714  0.002290
time           -0.013541   0.011958  -1.132   0.2574  -0.036978  0.009895
_cons           2.771597   0.160581  17.260   0.0000   2.456863  3.086330
log_sd(resid.) -0.383205   0.028194 -13.592   0.0000  -0.438465 -0.327946
\end{CodeOutput}
\end{CodeChunk}

The constant term for log valve gradient is estimated to be 2.772 (95\%
CI 2.457, 3.086) and this is estimated to change by -0.014 (95\% CI
-0.037, 0.010) for every year after valve replacement. The residual
error is reported in the results table as the log of the standard
deviation, meaning the residual standard error in this model is 0.682.
To assess model fit we can calculate the residuals using the
\code{predict} function to get the fitted values.

\begin{CodeChunk}

\begin{CodeInput}
R> heart.valve$m1res <- heart.valve$log.grad -  predict(m1, stat = "mu")
R>
R> library(ggplot2)
R> ggplot(heart.valve, aes(x = time, y = m1res)) +
R+   geom_point() +
R+   geom_hline(yintercept = 0, color = "blue") +
R+   xlab("Time (years)") +
R+   ylab("Residual") +
R+   theme_classic()
\end{CodeInput}
\begin{figure}

{\centering \includegraphics{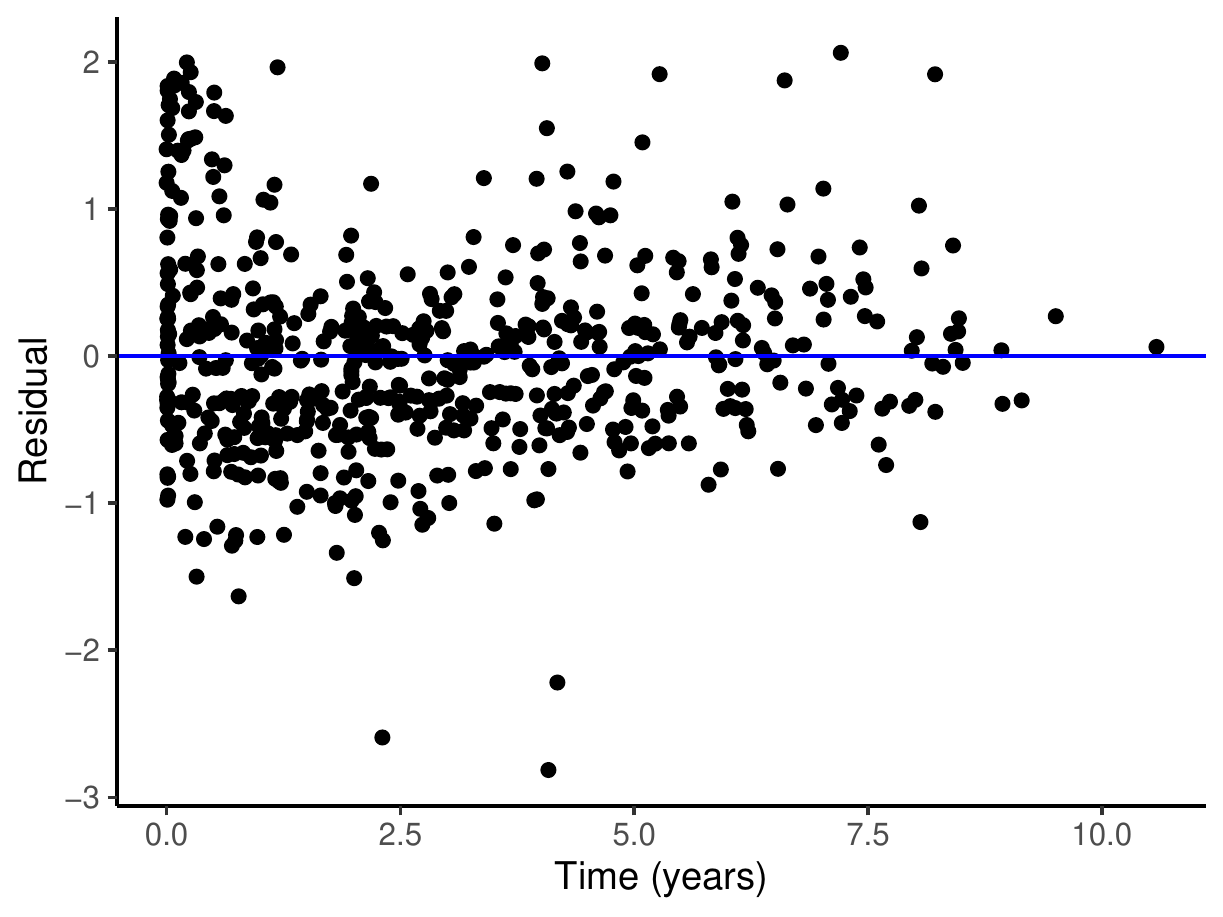}

}

\caption[Absolute residuals plot against time for linear model m1]{Absolute residuals plot against time for linear model m1}\label{fig:unnamed-chunk-6}
\end{figure}
\end{CodeChunk}

This shows that there seems to be some model misspecification, as the
values at the beginning and end are generally under predicted, while
values between 1 and 4 years are over predicted. To address this we can
add further flexibility to the shape of the log valve gradient over
time, using restricted cubic splines, with number degrees of freedom
specified as below. The boundary knots will be assumed to be at the
minimum and maximum of \code{log.grad} with the internal knots at
equally spaced centiles. Alternatively the locations of the knots can be
specified by using the \code{knots()} option. The spline terms have been
orthogonalised, which will impact on the interpretation of the intercept
term. While the spline terms themselves have little meaningful
interpretation they are reported to allow the model to be used to make
external predictions.

\begin{CodeChunk}

\begin{CodeInput}
R> m2 <- merlin(
R+   model = log.grad ~ sex + age + rcs(time, df = 3, orthog = TRUE),
R+   timevar = "time",
R+   family = "gaussian",
R+   data = heart.valve
R+ )
R> summary(m2)
\end{CodeInput}

\begin{CodeOutput}
Mixed effects regression model
Log likelihood = -678.6148

                Estimate Std. Error       z Pr(>|z|) [95% Conf. Interval]
sex            -0.090970   0.057917  -1.571  0.11625  -0.204486  0.022545
age             0.016740   0.002454   6.822  0.00000   0.011931  0.021550
rcs():1        -0.067805   0.026277  -2.580  0.00987  -0.119306 -0.016304
rcs():2        -0.215318   0.025901  -8.313  0.00000  -0.266084 -0.164553
rcs():3         0.097600   0.025675   3.801  0.00014   0.047278  0.147922
_cons           1.532048   0.159002   9.635  0.00000   1.220411  1.843685
log_sd(resid.) -0.448594   0.030335 -14.788  0.00000  -0.508050 -0.389139
\end{CodeOutput}
\end{CodeChunk}

When we plot the residuals for model \code{m2} we can see there is less
of a pattern over time, suggesting this model is a better fit to the
data.

\begin{CodeChunk}

\begin{CodeInput}
R> heart.valve$m2res <- heart.valve$log.grad -  predict(m2, stat = "mu")
R> ggplot(heart.valve, aes(x = time, y = m2res)) +
R+   geom_point() +
R+   geom_hline(yintercept = 0, color = "blue") +
R+   xlab("Time (years)") +
R+   ylab("Residual") +
R+   theme_classic()
\end{CodeInput}
\begin{figure}

{\centering \includegraphics{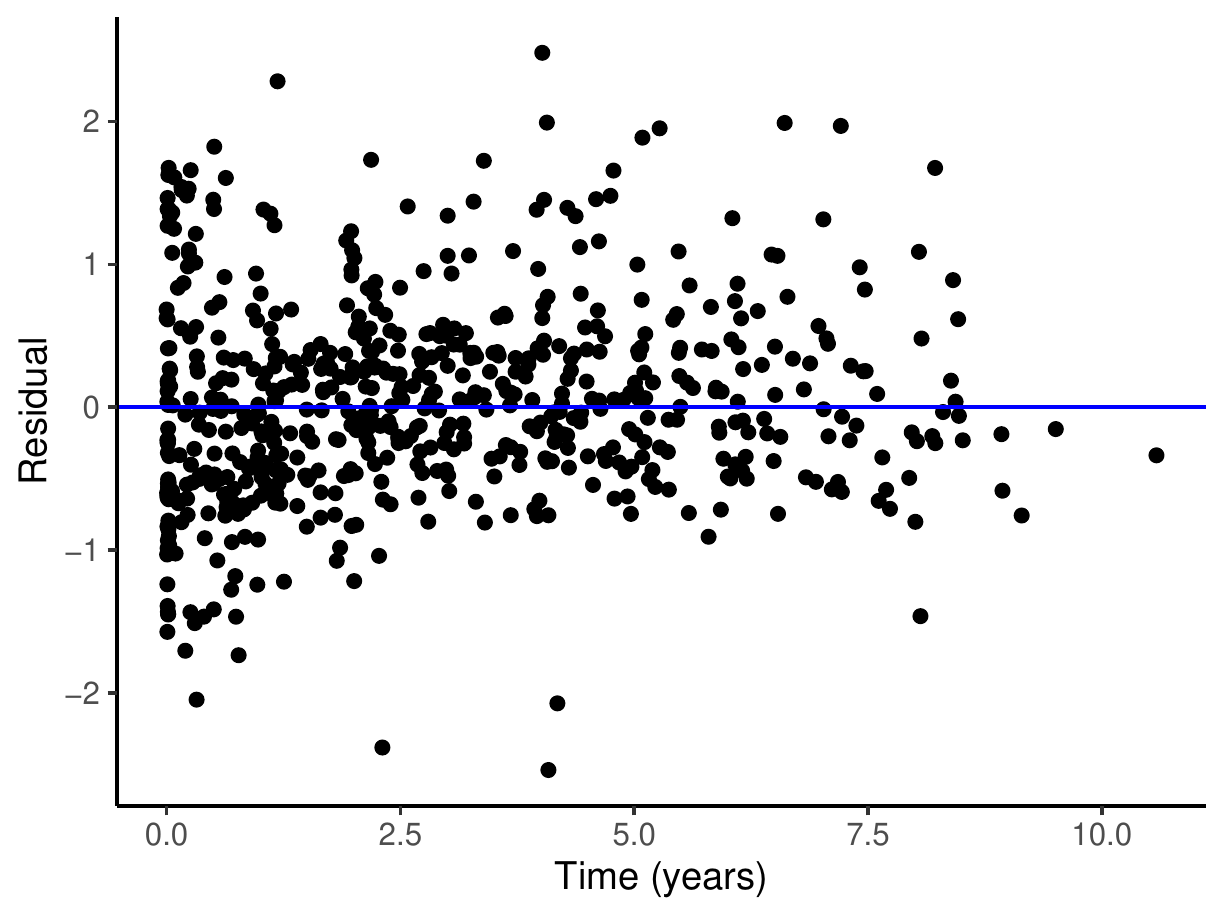}

}

\caption[Absolute residuals plot against time for restricted cubic spline model m2]{Absolute residuals plot against time for restricted cubic spline model m2}\label{fig:unnamed-chunk-7}
\end{figure}
\end{CodeChunk}

We can further improve this model by accounting for the clustered nature
of the \code{log.grad} measurements within patients (\code{id}). We can
add a normally-distributed random intercept at the patient \code{id}
level using the \code{M\#} syntax below. Each random effect is given a
name of this form to enable the sharing of random effects between
models, which will be illustrated later. In the model below \code{M1}
specifies a random intercept and \code{M2} specifies a random linear
slope. By default the random-effects at each level are not assumed to be
correlated (option \code{covariance(identity)}), however this can be
relaxed and the correlation estimated by instead specifying
\code{covariance(unstructured)}. For mixed-effects models the levels
must be specified using the \code{level} option. There is no limit to
the number of levels which can be fitted, but the levels must be be
specified from highest to lowest, e.g.~county \textgreater{} practice
\textgreater{} patient. By default all components in the model will have
an estimated coefficient, however the coefficient can be constrained to
1 using \code{*1} notation, which would normally be the case for random
effects not shared between models. By default estimation of the
likelihood is done using Gauss-Hermite quadrature with 7 nodes,
increasing this number using the \code{ip} option will improve
estimation of the likelihood, although this will increase computation
time considerably.

\begin{CodeChunk}

\begin{CodeInput}
R> m3 <- merlin(
R+   model = log.grad ~ sex + age + rcs(time, df = 3, orthog = TRUE) +
R+     M1[id] * 1 + time:M2[id] * 1,
R+   timevar = "time",
R+   level = "id",
R+   covariance = "unstructured",
R+   family = "gaussian",
R+   data = heart.valve
R+ )
R> summary(m3)
\end{CodeInput}

\begin{CodeOutput}
Mixed effects regression model
Log likelihood = -612.5806

                    Estimate Std. Error       z Pr(>|z|) [95% Conf.  Interval]
sex                0.1559667  0.0741357   2.104   0.0354  0.0106635  0.3012699
age                0.0057777  0.0030691   1.883   0.0598 -0.0002377  0.0117931
rcs():1           -0.0331820  0.0332446  -0.998   0.3182 -0.0983402  0.0319762
rcs():2           -0.1828734  0.0262125  -6.977   0.0000 -0.2342490 -0.1314978
rcs():3            0.1294719  0.0233294   5.550   0.0000  0.0837471  0.1751967
_cons              2.2338384  0.2030426  11.002   0.0000  1.8358823  2.6317945
log_sd(resid.)    -0.7233108  0.0342331 -21.129   0.0000 -0.7904063 -0.6562152
log_sd(M1)        -1.0672644  0.1262396  -8.454   0.0000 -1.3146895 -0.8198394
log_sd(M2)        -0.7900452  0.1528337  -5.169   0.0000 -1.0895939 -0.4904966
atanh_corr(M2,M1) -2.1512169  0.1571694 -13.687   0.0000 -2.4592632 -1.8431706

Integration method: Non-adaptive Gauss-Hermite quadrature
Integration points: 7
\end{CodeOutput}
\end{CodeChunk}

Adding the random-effects terms at the \code{id} level greatly reduces
the log-likelihood. The standard deviation for the random intercept is
0.344, and for the random slope is 0.454. The correlation between these
random-effects is reported as the inverse hyperbolic tangent, to get the
estimate of the correlation the \code{tanh()} function can be used,
\code{tanh(-2.151) = -0.973} showing that these random-effects are
highly correlated. Random-effects at multiple hierarchical levels can be
included by changing the level variable in square brackets, and by
specifying the levels from highest to lowest in the \code{level} option.

We can make predictions of the expected value of the response from
mixed-effects model \code{m3} using the \code{predict} function with the
\code{mu} option. Predictions will only be made for non-missing values
of the response. These predictions are marginal, calculated by
integrating out the random-effects, giving population averaged
predictions.

\begin{CodeChunk}

\begin{CodeInput}
R> ldata <- heart.valve[!is.na(heart.valve$log.grad), ]
R> ldata$pred1 <- predict(m3, stat = "mu", predmodel = 1, type = "marginal")
R> print(
R+   ldata[ldata$id %in% c(1:2, 13), c("id", "time", "log.grad", "pred1")],
R+   row.names = FALSE
R+ )
\end{CodeInput}

\begin{CodeOutput}
 id      time log.grad    pred1
  1 0.0109589 2.302585 3.122391
  1 3.6794520 2.302585 2.523278
  1 4.6958900 2.302585 2.645854
  2 6.3643840 2.639057 2.603468
  2 7.3041100 2.197225 2.642733
  2 8.3013700 2.484907 2.665921
 13 0.1369863 2.708050 3.177508
 13 1.0575340 2.833213 2.720037
 13 2.0547950 2.995732 2.514966
 13 3.9726030 3.401197 2.689035
\end{CodeOutput}
\end{CodeChunk}

\hypertarget{user-defined-model}{%
\subsection{User-defined model}\label{user-defined-model}}

As well as a wide range of standard models, \pkg{merlin} also allows
users the flexibility to specify their own likelihood functions using
the \code{user} family.

To help users to define their own likelihood there are a number of
inbuilt utility functions.

\begin{itemize}
\tightlist
\item
  \code{merlin_util_depvar(M)} - returns the dependent variable for the
  current model. For time-to-event outcomes this will be a matrix with
  two columns, for event time and event-indicator.
\item
  \code{merlin_util_xzb(M, t)} - returns the complex predictor for the
  current model, optionally evaluated at time \code{t}.
\item
  \code{merlin_util_xzb_deriv(M, t)} - returns the derivative with
  respect to time of the complex linear predictor for the current model,
  optionally evaluated at time \code{t}.
\item
  \code{merlin_util_xzb_deriv2(M, t)} - returns the second derivative
  with respect to time of the complex linear predictor for the current
  model, optionally evaluated at time \code{t}.
\item
  \code{merlin_util_xzb_integ(M, t)} - returns the integral with respect
  to time of the complex linear predictor for model \code{M}, optionally
  evaluated at time \code{t}.
\item
  \code{merlin_util_expval(M, t)} - returns the expected value of the
  response for the current model, optionally evaluated at time \code{t}.
\item
  \code{merlin_util_expval_deriv(M, t)} - returns the derivative with
  respect to time of the expected value of the response for the current
  model, optionally evaluated at time \code{t}.
\item
  \code{merlin_util_expval_deriv2(M, t)} - returns the second derivative
  with respect to time of the expected value of the response for the
  current model, optionally evaluated at time \code{t}.
\item
  \code{merlin_util_expval_integ(M, t)} - returns the integral with
  respect to time of the expected value of the response for the current
  model,, optionally evaluated at time \code{t}.
\item
  \code{merlin_util_ap(M,i)} - returns the \code{i}th ancillary
  parameter of the current model.
\item
  \code{merlin_util_timevar(M)} - returns the time variable for he
  current model, specified by the \code{timevar} option.
\end{itemize}

These utility functions take a list as input, which has been referred to
as \code{gml} below. This contains a \pkg{merlin} object, which should
not then be edited by the user. The \code{xzb} or \code{expval}
functions have a corresponding \code{*_mod()} function, which allows
users to specify an additional argument for which model to call,
e.g.~\code{merlin_util_xzb_mod(M,2)} will return the complex predictor
for the second model in the \code{merlin} statement, allowing submodels
to be linked.

The log-likelihood is specified as a function, giving the observation
level log-likelihood contribution. As an example a simple linear model
can be fitted using the function below.

\begin{CodeChunk}

\begin{CodeInput}
R> logl_gaussian <- function(gml) {
R+   y <- merlin_util_depvar(gml)
R+   xzb <- merlin_util_xzb(gml)
R+   se <- exp(merlin_util_ap(gml, 1))
R+
R+   mu <- (sweep(xzb, 1, y, "-"))^2
R+   logl <- ((-0.5 * log(2 * pi) - log(se)) - (mu / (2 * se^2)))
R+   return(logl)
R+ }
\end{CodeInput}
\end{CodeChunk}

To specify a user defined function, the family is given as \code{user},
the \code{userf} option must then be given the function above.

\begin{CodeChunk}

\begin{CodeInput}
R> m4 <- merlin(log.grad ~ sex + age + time + ap(1),
R+   family = "user",
R+   userf = "logl_gaussian",
R+   data = heart.valve
R+ )
R> summary(m4)
\end{CodeInput}

\begin{CodeOutput}
Mixed effects regression model
Log likelihood = -651.4753

       Estimate Std. Error       z Pr(>|z|) [95% Conf. Interval]
sex    0.140489   0.059778   2.350   0.0188   0.023327  0.257651
age   -0.002212   0.002297  -0.963   0.3356  -0.006714  0.002290
time  -0.013541   0.011958  -1.132   0.2574  -0.036978  0.009895
_cons  2.771597   0.160581  17.260   0.0000   2.456863  3.086330
_ap1  -0.383205   0.028194 -13.592   0.0000  -0.438465 -0.327946
\end{CodeOutput}
\end{CodeChunk}

The parameter estimates from this model are the same as model \code{M1}
above, where \code{_ap1} is the ancillary residual error parameter.
These user defined functions allows users to extend \pkg{merlin}, which
is particularly useful for those doing methodological research.

\hypertarget{survival-time-to-event-analysis}{%
\subsection{Survival / time-to-event
analysis}\label{survival-time-to-event-analysis}}

A number of standard time-to-event models are available in \pkg{merlin}
such as Weibull, exponential and Gompertz models. Additionally a range
of more flexible models are also available including Royston-Parmar
models, and a model on the log hazard scale, for both a number of forms
can be used for the baseline including restricted cubic splines, or
fractional polynomials.

\hypertarget{weibull-proportional-hazards-model}{%
\subsubsection{Weibull proportional hazards
model}\label{weibull-proportional-hazards-model}}

We will start by fitting a simple Weibull proportional hazard model for
time-to-death, adjusting for age and type of aortic valve replacement.
In order to fit a survival model a \code{Surv} object must be supplied
with the time and event indicator variables.

\begin{CodeChunk}

\begin{CodeInput}
R> m5 <- merlin(
R+   model = Surv(stime, died) ~ age + type,
R+   family = "weibull",
R+   data = heart.valve
R+ )
R> summary(m5)
\end{CodeInput}

\begin{CodeOutput}
Mixed effects regression model
Log likelihood = -173.3746

            Estimate Std. Error      z Pr(>|z|) [95% Conf. Interval]
age          0.09731    0.01903  5.115   0.0000    0.06002   0.13460
type         0.03834    0.34428  0.111   0.9113   -0.63644   0.71312
_cons      -11.68669    1.49113 -7.837   0.0000  -14.60925  -8.76413
log(gamma)   0.64107    0.12323  5.202   0.0000    0.39954   0.88261
\end{CodeOutput}
\end{CodeChunk}

The results table gives the coefficient for the factors in the model,
which is the log of the hazard ratio. Therefore the hazard ratio for
type of aortic valve replacement is \code{exp(0.038) =} 1.039 showing
that having a stentless valve replacement leads to worse survival than
having a homograft valve replacement, although this is not statistically
significant.

The survival function can be obtained using the \code{predict} function,
with the \code{survival} option. Predictions will only be made for
non-missing values of the response.

\begin{CodeChunk}

\begin{CodeInput}
R> sdata <- heart.valve[!is.na(heart.valve$stime), ]
R> sdata$pred2 <- predict(m5, stat = "survival", predmodel = 1, type = "fixedonly")
R> print(
R+   sdata[
R+     sdata$id %in% c(1:2, 13),
R+     c("id", "stime", "died", "age", "type", "pred2")
R+   ],
R+   row.names = FALSE
R+ )
\end{CodeInput}

\begin{CodeOutput}
 id    stime died      age type     pred2
  1 4.956164    0 75.06027    1 0.7625097
  2 9.663014    0 45.79452    0 0.9476903
 13 5.186301    1 69.94247    0 0.8412621
\end{CodeOutput}
\end{CodeChunk}

The predictions give the survival probability for each individual at
their event time, depending on their age and type of valve replacement.
We can use the \code{at} option to compare the survival functions for
different levels of a covariate, while holding other covariates
constant. As type of valve replacement has a smaller effect, we will
instead look at the differences in survival prediction by age, assuming
a stentless value replacement.

\begin{CodeChunk}

\begin{CodeInput}
R> p_50 <- predict(m5, stat = "survival", type = "fixedonly",
R+                 at = c(age = 50, type = 1))
R> p_60 <- predict(m5, stat = "survival", type = "fixedonly",
R+                 at = c(age = 60, type = 1))
R> p_70 <- predict(m5, stat = "survival", type = "fixedonly",
R+                 at = c(age = 70, type = 1))
\end{CodeInput}
\end{CodeChunk}

\begin{CodeChunk}

\begin{CodeInput}
R> surv_pred <- data.frame(p_50, p_60, p_70,
R+                         stime = heart.valve[!is.na(heart.valve$stime), "stime"])
R> surv_pred <- surv_pred[!duplicated(surv_pred$stime),]
R> surv_pred <- melt(surv_pred, id.var = "stime")
R>
R> ggplot(surv_pred, aes(x = stime, y = value, linetype = variable)) +
R+   geom_line(size = 0.6) +
R+   coord_cartesian(ylim = c(0, 1)) +
R+   xlab("Time (years)") +
R+   ylab("Survival probability") +
R+   theme_classic() +
R+   theme(legend.position = c(0.2, 0.2)) +
R+   scale_linetype_discrete(
R+     name = "Age",
R+     labels = c("50 years", "60 years", "70 years"))
\end{CodeInput}
\begin{figure}

{\centering \includegraphics{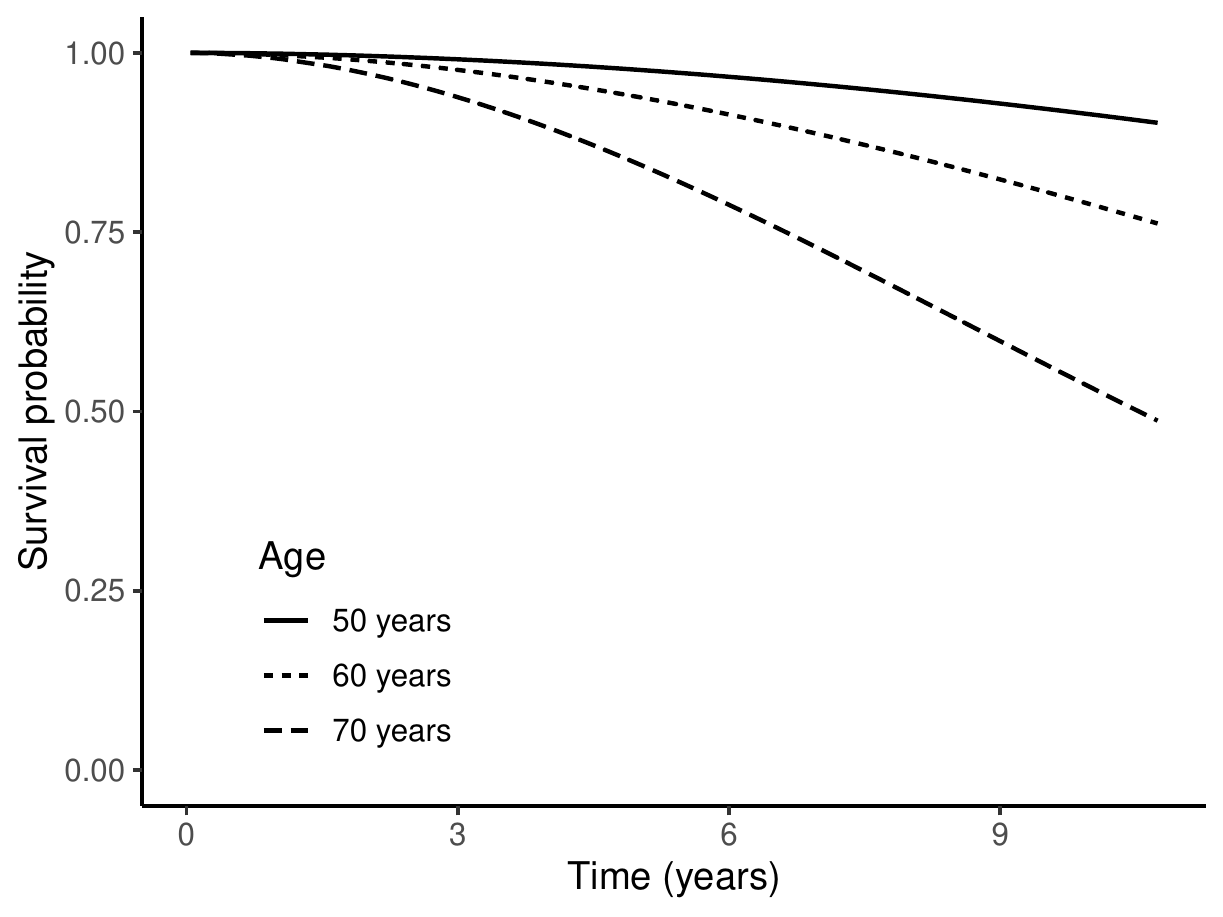}

}

\caption[Survival functions for patients following stentless value replacement by age (from model m5)]{Survival functions for patients following stentless value replacement by age (from model m5)}\label{fig:unnamed-chunk-12}
\end{figure}
\end{CodeChunk}

This gives a visual representation of how the survival probability
changes depending on age at entry, for a particular treatment option.

\hypertarget{spline-based-survival-model}{%
\subsubsection{Spline-based survival
model}\label{spline-based-survival-model}}

Further flexibility can be included in survival models by modelling the
hazard function using splines. Royston-Parmar models use restricted
cubic splines to model the baseline log cumulative hazard function. They
allow flexibility in the shape of the baseline hazard and allow for
time-dependent effects. The form of the baseline hazard is specified by
adding the function to the linear predictor of the survival model. Here
we fit a model using restricted cubic splines with 3 degrees of freedom
in the baseline hazard. Using the \code{event = TRUE} option means the
knots locations for the the splines are based on the event times only,
ignoring censored time points.

\begin{CodeChunk}

\begin{CodeInput}
R> m6 <- merlin(
R+   model = Surv(stime, died) ~ age + type +
R+     rcs(stime, df = 3, log = TRUE, event = TRUE),
R+   timevar = "stime",
R+   family = "rp",
R+   data = heart.valve
R+ )
R> summary(m6)
\end{CodeInput}

\begin{CodeOutput}
Mixed effects regression model
Log likelihood = -170.653

          Estimate Std. Error      z Pr(>|z|) [95% Conf.  Interval]
age       0.101765   0.019528  5.211   0.0000   0.063491   0.140040
type      0.025291   0.343717  0.074   0.9413  -0.648381   0.698964
rcs():1   1.107442   0.143163  7.736   0.0000   0.826849   1.388036
rcs():2  -0.274877   0.089043 -3.087   0.0020  -0.449398  -0.100357
rcs():3  -0.002397   0.076386 -0.031   0.9750  -0.152112   0.147317
_cons    -9.057037   1.401644 -6.462   0.0000 -11.804210  -6.309865
\end{CodeOutput}
\end{CodeChunk}

The estimated age and treatment effects are very similar to the Weibull
survival model \code{m5} above. We can compare the shape in baseline
hazards between the Weibul and Royston-Parmar using the \code{hazard}
option in the \code{predict} function.

\begin{CodeChunk}

\begin{CodeInput}
R> base_m5 <- predict(m5, stat = "hazard", type ="fixedonly",
R+                    at = c(age = 0, type = 0))
R> base_m6 <- predict(m6, stat = "hazard", type ="fixedonly",
R+                    at = c(age = 0, type = 0))
R>
R> base_pred <- data.frame(stime = heart.valve[!is.na(heart.valve$stime), "stime"],
R+                         base_m5,
R+                         base_m6)
R> base_pred <- base_pred[!duplicated(base_pred$stime),]
R> base_pred <- melt(base_pred, id.var = "stime")
\end{CodeInput}
\end{CodeChunk}

\begin{CodeChunk}

\begin{CodeInput}
R> ggplot(base_pred, aes(x = stime, y = value, linetype = variable)) +
R+   geom_line(size = 0.6) +
R+   xlab("Time (years)") + ylab("Baseline hazard") +
R+   theme_classic() + theme(legend.position = c(0.2, 0.8)) +
R+   scale_linetype_discrete(name = "Model", labels = c("Weibull", "RP - 3df"))
\end{CodeInput}
\begin{figure}

{\centering \includegraphics{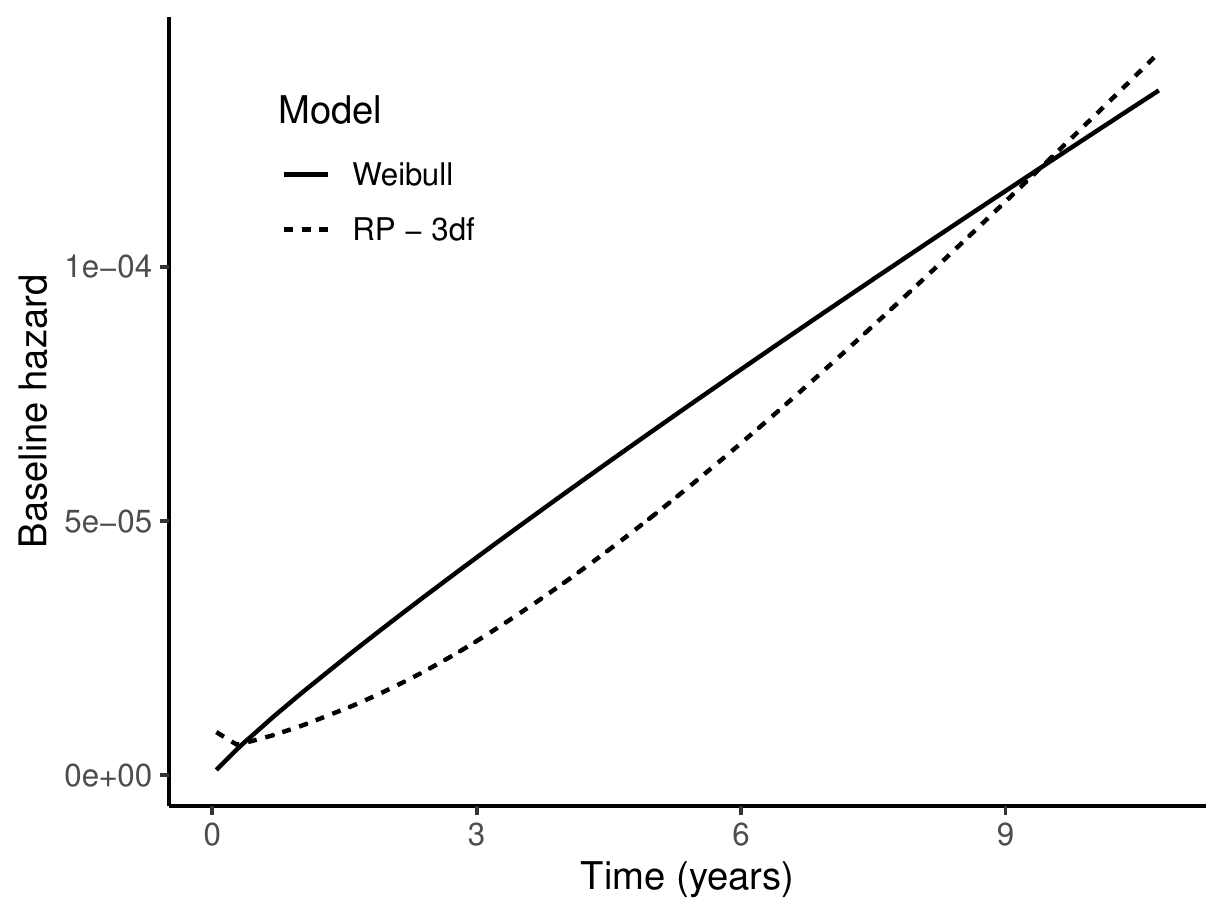}

}

\caption[Baseline hazard functions with Weibull model (m5) and Royston-Parmar model with 3 degrees of freedom (m6)]{Baseline hazard functions with Weibull model (m5) and Royston-Parmar model with 3 degrees of freedom (m6)}\label{fig:unnamed-chunk-14}
\end{figure}
\end{CodeChunk}

\hypertarget{assess-proportional-hazards}{%
\subsubsection{Assess
proportional-hazards}\label{assess-proportional-hazards}}

The survival models above assume proportional hazards. We can test this
assumption in the effect of type aortic valve replacement by including
the interaction between type and log time. It is important to use the
\code{timevar} option for this model, as time-dependent effects need to
be differentiated with respect to time to calculate the hazard function.

\begin{CodeChunk}

\begin{CodeInput}
R> m7 <- merlin(
R+   model = Surv(stime, died) ~ age + type +
R+     type:fp(stime, powers = c(0)) +
R+     rcs(stime, df = 3, log = TRUE, event = TRUE),
R+   timevar = "stime",
R+   family = "rp",
R+   data = heart.valve
R+ )
R> summary(m7)
\end{CodeInput}

\begin{CodeOutput}
Mixed effects regression model
Log likelihood = -170.3792

           Estimate Std. Error      z Pr(>|z|) [95% Conf. Interval]
age         0.10223    0.01956  5.226   0.0000    0.06389   0.14057
type       -0.62548    0.93454 -0.669   0.5033   -2.45714   1.20618
type:fp()   0.34704    0.46673  0.744   0.4571   -0.56773   1.26180
rcs():1     0.96557    0.22712  4.251   0.0000    0.52044   1.41071
rcs():2    -0.24625    0.09397 -2.620   0.0088   -0.43044  -0.06207
rcs():3    -0.01036    0.07610 -0.136   0.8917   -0.15952   0.13880
_cons      -9.01176    1.40479 -6.415   0.0000  -11.76510  -6.25841
\end{CodeOutput}
\end{CodeChunk}

The interaction term (\code{type:fp()}) is not significant, therefore
accounting for time-dependent effects on the type of aortic valve is not
necessary.

\hypertarget{non-linear-effects}{%
\subsubsection{Non-linear effects}\label{non-linear-effects}}

In the above models the effect of age was assumed to be linear. We can
investigate the non-linear effect of age using fractional polynomials.
Including \code{fp(age, powers = c(1, 1))} specifies a second-order
fractional polynomial. The first specified term is \code{stime} to the
first power, which in this case is a linear term. The second specified
term is \code{stime} to the second power multiplied by the natural log
of \code{stime}.

\begin{CodeChunk}

\begin{CodeInput}
R> m8 <- merlin(
R+   model = Surv(stime, died) ~ type +
R+     fp(age, powers = c(1, 1)) +
R+     rcs(stime, df = 3, log = TRUE, event = TRUE),
R+   timevar = "stime",
R+   family = "rp",
R+   data = heart.valve
R+ )
R> summary(m8)
\end{CodeInput}

\begin{CodeOutput}
Mixed effects regression model
Log likelihood = -170.7799

          Estimate Std. Error      z Pr(>|z|) [95% Conf.  Interval]
type      0.038219   0.350542  0.109   0.9132  -0.648830   0.725268
fp():1   -0.596800   0.392650 -1.520   0.1285  -1.366380   0.172780
fp():2    0.133596   0.072833  1.834   0.0666  -0.009154   0.276345
rcs():1   1.114711   0.143890  7.747   0.0000   0.832692   1.396731
rcs():2  -0.277742   0.089596 -3.100   0.0019  -0.453347  -0.102137
rcs():3  -0.003194   0.076810 -0.042   0.9668  -0.153740   0.147351
_cons     0.007325   5.537987  0.001   0.9989 -10.846930  10.861580
\end{CodeOutput}
\end{CodeChunk}

The hazard ratios for linear age (0.551) and linear age multiplied by
log age (1.143) are both significant, suggesting the effect of age is
non-linear. We could go on to investigate whether the proportional
hazard assumption is valid for this non-linear function of age by
fitting an interaction between the age function and log time.

\hypertarget{wrapper}{%
\subsection{Wrapper functions}\label{wrapper}}

The syntax for relatively simple one outcome models such as those above
can be simplified using available wrapper functions, which use the
powerful \pkg{merlin} function underneath. For example the \code{mlsurv}
wrapper fits parametric survival models, with \code{exponential},
\code{weibull}, \code{gompertz}, \code{rp}, \code{logchazard}, and
\code{loghazard} model options. To illustrate, the Weibull model in
\code{m5} above can fitted using the wrapper \code{mlsurv}, with
simplified syntax.

\begin{CodeChunk}

\begin{CodeInput}
R> mlsurv(
R+   formula = Surv(stime, died) ~ age + type,
R+   distribution = "weibull",
R+   data = heart.valve
R+ )
\end{CodeInput}

\begin{CodeOutput}
Proportional hazards regression model
Weibull baseline hazard
Data: data

Coefficients:
       age        type       _cons  log(gamma)
   0.09719     0.03778   -11.67541     0.63998
\end{CodeOutput}
\end{CodeChunk}

\hypertarget{competing-risks}{%
\subsection{Competing risks}\label{competing-risks}}

Competing risk analysis can be framed as a multiple outcome survival
model by specifying cause-specific hazard models. As this data set only
contains all cause survival information, to illustrate fitting a
competing risks model we randomly assign the deaths to either
cardiovascular disease (\code{cardio}) or other causes (\code{other}).

\begin{CodeChunk}

\begin{CodeInput}
R> set.seed(6342)
R> heart.valve$cardio[!is.na(heart.valve$died)] <-
R+   rbinom(length(heart.valve$died[!is.na(heart.valve$died)]), 1, 0.6)
R> heart.valve$other <- 1 - heart.valve$cardio
R> heart.valve$cardio[heart.valve$died == 0 & !is.na(heart.valve$died)] <- 0
R> heart.valve$other[heart.valve$died == 0 & !is.na(heart.valve$died)] <- 0
\end{CodeInput}
\end{CodeChunk}

We can then fit a model with \texttt{cardio} as one outcome and
\texttt{other} as a second outcome. Both event types have been fitted
using a Royston-Parmar model with 3 degrees of freedom, however it is
possible to use different survival model types for each event. As there
are two outcomes the \code{model} now needs to be specified as a list.

\begin{CodeChunk}

\begin{CodeInput}
R> m9 <- merlin(
R+   model = list(
R+     Surv(stime, cardio) ~ type +
R+       rcs(stime, df = 3, log = TRUE, event = TRUE),
R+     Surv(stime, other) ~ type +
R+       rcs(stime, df = 3, log = TRUE, event = TRUE)
R+   ),
R+   timevar = c("stime", "stime"),
R+   family = c("rp", "rp"),
R+   data = heart.valve
R+ )
R> summary(m9)
\end{CodeInput}

\begin{CodeOutput}
Mixed effects regression model
Log likelihood = -232.9797

         Estimate Std. Error      z Pr(>|z|) [95% Conf. Interval]
type    -0.004186   0.412139 -0.010   0.9919  -0.811963  0.803591
rcs():1  1.098643   0.186195  5.901   0.0000   0.733708  1.463578
rcs():2 -0.415694   0.101863 -4.081   0.0000  -0.615343 -0.216045
rcs():3 -0.149017   0.100844 -1.478   0.1395  -0.346668  0.048634
_cons   -2.836043   0.351496 -8.068   0.0000  -3.524962 -2.147124
type    -0.408162   0.364551 -1.120   0.2629  -1.122670  0.306346
rcs():1  2.480418   0.792595  3.129   0.0018   0.926960  4.033876
rcs():2  1.362926   0.631493  2.158   0.0309   0.125223  2.600629
rcs():3 -0.178580   0.124901 -1.430   0.1528  -0.423381  0.066222
_cons   -2.405594   0.323577 -7.434   0.0000  -3.039795 -1.771394
\end{CodeOutput}
\end{CodeChunk}

The hazard ratio for type of graft is 0.996 for death from
cardiovascular disease, suggesting in this simulated example having a
stentless valve replacement increases the risk of death due to
cardiovascular disease compared to having a homograft valve replacement.
The effect is in the same direction for death from other causes,
although the hazard ratio of 0.665 suggests the effect is much smaller.

Hazard ratios describe the relative differences in hazard between
groups. In the case of competing risks the cause-specific cumulative
incidence function, which is the probability of failure from the event
of interest in the presence of other competing events, may be more
useful. We can calculate the cause-specific cumulative incidence
function for each of the causes in the model using the \code{predict}
function, by specifying which submodel to predict from using the
\code{predmodel} option. Predictions are for each valve type at a given
age, specified using the \code{at} option. Here the new
\code{cardio_homograft} variable will give the cause-specific cumulative
incidence function for time-to-death from cardiovascular disease (which
is submodel 1) assuming a patient aged 50 has had a homograft valve
replacement.

\begin{CodeChunk}

\begin{CodeInput}
R> card_homo    <- predict(m9, stat = "cif", type = "fixedonly",
R+                        predmodel = 1, at = c(age = 50, type = 0))
\end{CodeInput}
\end{CodeChunk}

To create stacked cumulative incidence plots for both types of value
replacements we then calculate further predictions as below for
stentless grafts, and for death from other causes.

\begin{CodeChunk}

\begin{CodeInput}
R> card_stent  <- predict(m9, stat = "cif", type = "fixedonly",
R+                             predmodel = 1, at = c(age = 50, type = 1))
R> other_homo  <- predict(m9, stat = "cif", type = "fixedonly",
R+                             predmodel = 2, at = c(age = 50, type = 0))
R> other_stent <- predict(m9, stat = "cif", type = "fixedonly",
R+                             predmodel = 2, at = c(age = 50, type = 1))
\end{CodeInput}
\end{CodeChunk}

We can then plot the stacked cumulative incidence plots, allowing us to
compare between the two types of valve replacement. They show that for
both types the cumulative incidence for cardiovascular disease starts to
flatten over time, whereas it continues to increase for death from other
causes.

\begin{CodeChunk}

\begin{CodeInput}
R> stime = rep(heart.valve$stime[!is.na(heart.valve$stime)], 4)
R> pred  = c(card_homo, card_stent, other_homo, other_stent)
R> valve = rep(c(rep("Homograft valve", length(card_homo)),
R+               rep("Stentless valve", length(card_stent))), 2)
R> cod   = c(rep("cardio", length(card_homo) * 2),
R+           rep("other", length(card_stent) * 2))
R> pred_comp <- data.frame(stime, pred, valve, cod)
R> pred_comp <- pred_comp[!duplicated(pred_comp[, c(1, 3, 4)]),]
\end{CodeInput}
\end{CodeChunk}

\begin{CodeChunk}

\begin{CodeInput}
R> ggplot(pred_comp, aes(x = stime, y = pred, fill = cod)) +
R+   geom_area() +
R+   xlab("Time (years)") +
R+   ylab("Cumulative Incidence") +
R+   theme_classic(base_size = 10) +
R+   theme(legend.position = c(0.19, 0.83)) +
R+   scale_fill_discrete(name = "Cause of death",
R+                       labels = c("Other", "Cardiovascular disease"),
R+                       guide = guide_legend(reverse = TRUE)) +
R+   facet_grid(. ~ valve)
\end{CodeInput}
\begin{figure}

{\centering \includegraphics{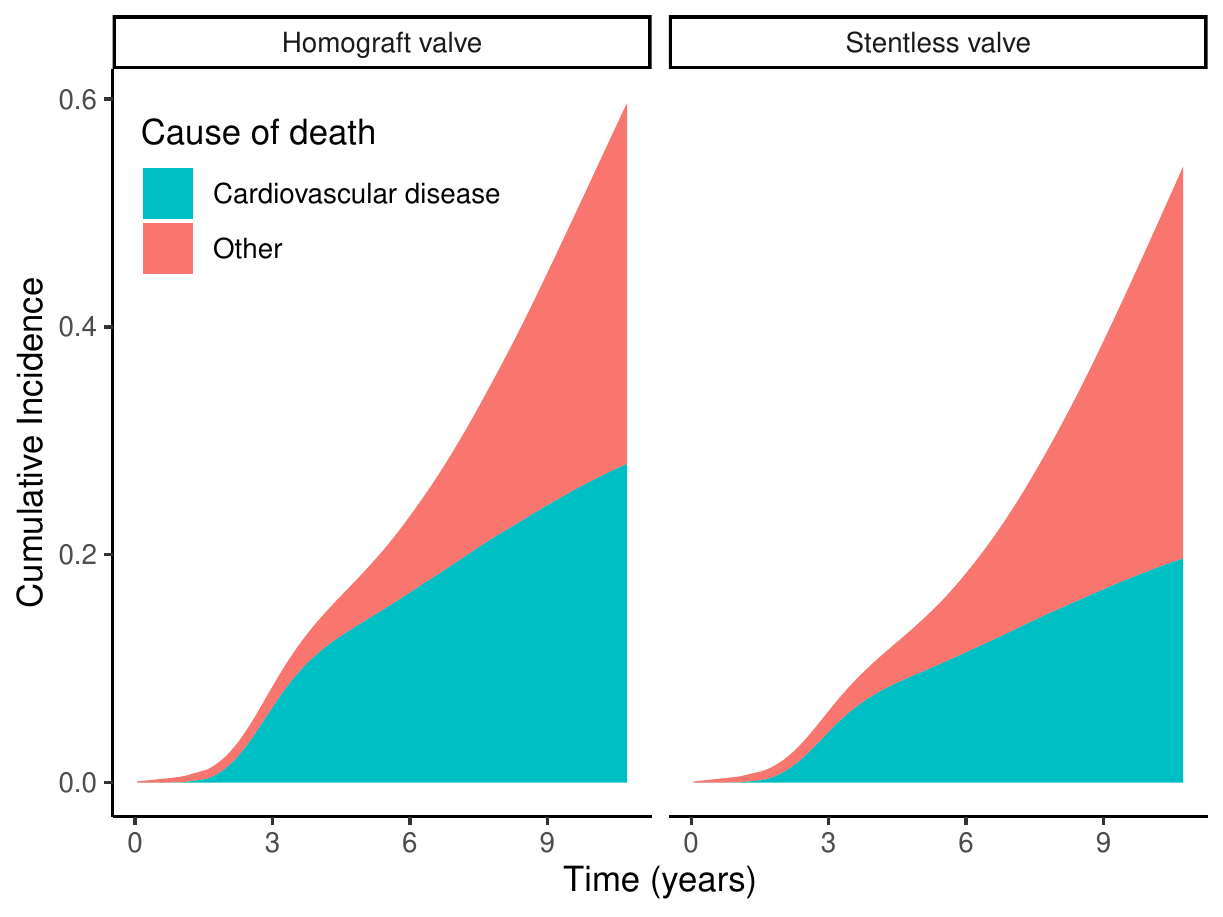}

}

\caption[Stacked cumulative incidence functions for death from cardiovascular disease and death from other causes, by type of valve replacement (model m9)]{Stacked cumulative incidence functions for death from cardiovascular disease and death from other causes, by type of valve replacement (model m9)}\label{fig:unnamed-chunk-20}
\end{figure}
\end{CodeChunk}

\hypertarget{multiple-outcome-models}{%
\subsection{Multiple-outcome models}\label{multiple-outcome-models}}

We can also use \pkg{merlin} to fit joint longitudinal survival models.
This can firstly be done through shared random-effects. In \pkg{merlin}
shared random-effects are specified by using the same name for the
random-effects that are to be shared. When using shared-random effects
it is possible to allow for an association factor, which explains the
relationship between the individual-specific random intercept of the
biomarker and their survival. To estimate the association parameter the
\code{*1} is dropped from one of the random-effects terms, removing the
constraint and allowing a coefficient to be estimated.

\begin{CodeChunk}

\begin{CodeInput}
R> m10 <- merlin(
R+   model = list(
R+     Surv(stime, died) ~ type + M1[id],
R+     log.grad ~ sex + age + time + M1[id] * 1
R+   ),
R+   timevar = c("stime", "time"),
R+   levels = c("id"),
R+   family = c("weibull", "gaussian"),
R+   data = heart.valve
R+ )
R> summary(m10)
\end{CodeInput}

\begin{CodeOutput}
Mixed effects regression model
Log likelihood = -825.2692

                 Estimate Std. Error       z Pr(>|z|) [95% Conf.  Interval]
type            0.6198651  0.3254017   1.905   0.0568 -0.0179105  1.2576408
M1              1.8384968  0.8677693   2.119   0.0341  0.1377002  3.5392934
_cons          -5.0586043  0.5317924  -9.512   0.0000 -6.1008982 -4.0163104
log(gamma)      0.5320193  0.1265303   4.205   0.0000  0.2840245  0.7800141
sex             0.1466814  0.0745633   1.967   0.0492  0.0005401  0.2928228
age            -0.0051930  0.0030689  -1.692   0.0906 -0.0112078  0.0008219
time           -0.0017929  0.0124788  -0.144   0.8858 -0.0262510  0.0226652
_cons           2.9499413  0.2081879  14.170   0.0000  2.5419006  3.3579820
log_sd(resid.) -0.5013162  0.0340641 -14.717   0.0000 -0.5680806 -0.4345517
log_sd(M1)     -1.0821441  0.1140678  -9.487   0.0000 -1.3057129 -0.8585752

Integration method: Non-adaptive Gauss-Hermite quadrature
Integration points: 7
\end{CodeOutput}
\end{CodeChunk}

The parameter estimates from the survival submodel are reported first,
followed by the estimates from the longitudinal model, then the
random-effect estimates. Here the random effect \code{M1} is shared
between the longitudinal and survival submodels. In the longitudinal
model \code{M1} is a random intercept, therefore this model is relating
an individuals baseline valve gradient to their survival. The standard
deviation for the random effect is 0.339. The association between the
random-effect on the intercept and survival is given in row \code{M1} in
the results table, a value of 1.838 suggests that higher baseline valve
gradient leads to worse survival.

Alternatively it is possible to to use current value parameterisation,
linking the time-dependent expected value of the biomarker to survival
using the \code{EV[]} option. In order to do this the \code{timevar()}
must be supplied, as integration over time is required for the survival
model likelihood contribution.

\begin{CodeChunk}

\begin{CodeInput}
R> m11 <- merlin(
R+   model = list(
R+     Surv(stime, died) ~ type + EV[log.grad],
R+     log.grad ~ sex + age + time + M1[id] * 1
R+   ),
R+   timevar = c("stime", "time"),
R+   levels = c("id"),
R+   family = c("weibull", "gaussian"),
R+   data = heart.valve
R+ )
R> summary(m11)
\end{CodeInput}

\begin{CodeOutput}
Mixed effects regression model
Log likelihood = -826.1898

                 Estimate Std. Error       z Pr(>|z|) [95% Conf.  Interval]
type             0.814597   0.322574   2.525   0.0116   0.182363   1.446831
EV[]             0.785010   0.810309   0.969   0.3327  -0.803168   2.373187
_cons           -6.747273   2.315188  -2.914   0.0036 -11.284959  -2.209587
log(gamma)       0.403767   0.131396   3.073   0.0021   0.146236   0.661298
sex              0.174081   0.074095   2.349   0.0188   0.028858   0.319305
age             -0.003442   0.003240  -1.062   0.2882  -0.009793   0.002909
time            -0.011135   0.013790  -0.807   0.4194  -0.038163   0.015893
_cons            2.856742   0.226865  12.592   0.0000   2.412094   3.301390
log_sd(resid.)  -0.512929   0.033856 -15.150   0.0000  -0.579286  -0.446572
log_sd(M1)      -1.106644   0.114632  -9.654   0.0000  -1.331319  -0.881969

Integration method: Non-adaptive Gauss-Hermite quadrature
Integration points: 7
\end{CodeOutput}
\end{CodeChunk}

Now the association term \code{EV[]} describes the effect of the current
expected value of the biomarker \code{log.grad} on survival. The hazard
ratio is 2.192, meaning for every unit increase in log valve gradient
the hazard is 2.192 times higher.

The form of the association between the the longitudinal biomarker model
and survival is flexible. For example, if there was instead a random
slope term we could link the trend in the repeatedly measured biomarker
to survival by using the gradient of the longitudinal model
\code{dEV[]}.

\begin{CodeChunk}

\begin{CodeInput}
R> m12 <- merlin(
R+   model = list(
R+     Surv(stime, died) ~ type + dEV[log.grad],
R+     log.grad ~ sex + age + time + time:M1[id] * 1
R+   ),
R+   timevar = c("stime", "time"),
R+   levels = c("id"),
R+   family = c("weibull", "gaussian"),
R+   data = heart.valve
R+ )
R> summary(m12)
\end{CodeInput}

\begin{CodeOutput}
Mixed effects regression model
Log likelihood = -840.1187

                 Estimate Std. Error       z Pr(>|z|) [95% Conf.  Interval]
type             0.834296   0.331631   2.516   0.0119   0.184312   1.484281
dEV[]            1.835397  39.467308   0.047   0.9629 -75.519105  79.189898
_cons           -5.018370   0.719250  -6.977   0.0000  -6.428074  -3.608665
log(gamma)       0.517752   0.124721   4.151   0.0000   0.273304   0.762200
sex              0.141272   0.061993   2.279   0.0227   0.019767   0.262777
age             -0.001734   0.003074  -0.564   0.5727  -0.007758   0.004290
time            -0.017077   0.021456  -0.796   0.4261  -0.059131   0.024977
_cons            2.748958   0.198092  13.877   0.0000   2.360704   3.137211
log_sd(resid.)  -0.401866   0.034755 -11.563   0.0000  -0.469985  -0.333748
log_sd(M1)      -3.337529   0.660043  -5.057   0.0000  -4.631190  -2.043867

Integration method: Non-adaptive Gauss-Hermite quadrature
Integration points: 7
\end{CodeOutput}
\end{CodeChunk}

The hazard ratio for the gradient of the longitudinal model is 6.268
(\code{exp(dEV[]}), this large hazard ratio suggests an increase in log
of the valve gradient leads to greater hazard. However the confidence
intervals are wide, which may be due to the small standard deviation of
0.036 in the linear slope, leading to issues in the estimation of its
effect.

Other available links include the second derivative \code{d2EV[]} and
the cumulative exposure \code{iEV[]}. We can extend the previous model
to investigate possible non-linearities in the associations and time
dependent effects by including the interaction between the expected
value of the biomarker and log time.

\begin{CodeChunk}

\begin{CodeInput}
R> m13 <- merlin(
R+   model = list(
R+     Surv(stime, died) ~ type +
R+                         EV[log.grad] +
R+                         EV[log.grad]:fp(stime, powers = c(0)),
R+     log.grad ~ time + M1[id] * 1
R+   ),
R+   timevar = c("stime", "time"),
R+   levels = c("id"),
R+   family = c("weibull", "gaussian"),
R+   data = heart.valve
R+ )
R> summary(m13)
\end{CodeInput}

\begin{CodeOutput}
Mixed effects regression model
Log likelihood = -827.7168

                Estimate Std. Error       z Pr(>|z|) [95% Conf. Interval]
type             0.67433    0.31513   2.140   0.0324    0.05669   1.29197
EV[]             0.24692    1.03307   0.239   0.8111   -1.77786   2.27170
EV[]:fp()       -0.17456    0.38326  -0.455   0.6488   -0.92574   0.57661
_cons           -5.85705    3.03192  -1.932   0.0534  -11.79950   0.08539
log(gamma)       0.77284    0.46006   1.680   0.0930   -0.12885   1.67453
time            -0.01275    0.01273  -1.002   0.3163   -0.03770   0.01219
_cons            2.66026    0.04662  57.058   0.0000    2.56888   2.75164
log_sd(resid.)  -0.50952    0.03478 -14.652   0.0000   -0.57768  -0.44136
log_sd(M1)      -1.13669    0.12899  -8.812   0.0000   -1.38949  -0.88388

Integration method: Non-adaptive Gauss-Hermite quadrature
Integration points: 7
\end{CodeOutput}
\end{CodeChunk}

The parameter estimate for the interaction (\code{EV[]:fp()} in the
results table) suggests that over time a unit increase in the current
value of the log valve gradient has a reduced effect on hazard, however
this is not significant.

\hypertarget{a-final-model}{%
\subsection{A final model}\label{a-final-model}}

For illustrative purposes we will now show how flexible \pkg{merlin} is
by bringing together the previous examples, with some new model options,
in one final model. Binary outcomes can be included using the
\code{bernoulli} family. To illustrate this we will create a new binary
variable \code{catef} from the ejection fraction variable \code{ef}.

\begin{CodeChunk}

\begin{CodeInput}
R> heart.valve$catef <- 0
R> heart.valve$catef[heart.valve$ef > 70] <- 1
\end{CodeInput}
\end{CodeChunk}

Model \code{m14} includes the two survival models for the competing
risks of death (cardiovascular disease and other causes) from model
\code{m9}. Continuous log valve gradient over time is described using
restricted cubic splines as in model \code{m6} with random intercept
term (\code{M1}). Binary \code{catef} over time is also modelled with a
random intercept term (\code{M2}). Both survival outcomes are described
using Weibull models, the effect of the type of valve replacement on
both causes of death is estimated. The effect of type of valve
replacement is assumed to be time dependent in the time to death from
other causes model. The random intercept for valve gradient \code{M1} is
shared with the time-to-death from other causes model, while the random
intercept for categorical ejection fraction \code{M2} is shared with the
time-to-death from cardiovascular disease model. The expected value of
valve gradient is included in the time-to-death from cardiovascular
disease model.

\begin{CodeChunk}

\begin{CodeInput}
R> m14 <-  merlin(
R+   model = list(
R+     Surv(stime, cardio) ~ type + EV[log.grad] + M2[id],
R+     Surv(stime, other) ~ type + type:fp(stime, powers = c(0)) + M1[id],
R+     log.grad ~ age + type + rcs(time, df = 3, orthog = TRUE) + M1[id] * 1,
R+     catef ~ fp(time, powers = c(1)) + M2[id] * 1
R+   ),
R+   timevar = c("stime", "stime", "time", "time"),
R+   levels = c("id"),
R+   covariance = "unstructured",
R+   family = c("weibull", "weibull", "gaussian", "bernoulli"),
R+   data = heart.valve,
R+   control = list(ip = 9)
R+ )
R> summary(m14)
\end{CodeInput}

\begin{CodeOutput}
Mixed effects regression model
Log likelihood = -1409.455

                    Estimate Std. Error       z Pr(>|z|) [95% Conf.  Interval]
type                1.713733   0.741166   2.312   0.0208   0.261074   3.166392
EV[]               -1.206885   2.351814  -0.513   0.6078  -5.816356   3.402585
M2                 -0.069736   0.119881  -0.582   0.5608  -0.304699   0.165228
_cons              -8.339090   5.046409  -1.652   0.0984 -18.229870   1.551689
log(gamma)          1.429550   0.153848   9.292   0.0000   1.128013   1.731087
type               11.132882   1.326384   8.393   0.0000   8.533217  13.732548
type:fp()          -4.097624   0.684881  -5.983   0.0000  -5.439965  -2.755282
M1                  2.813494   0.158675  17.731   0.0000   2.502497   3.124491
_cons             -13.450441   1.494459  -9.000   0.0000 -16.379527 -10.521355
log(gamma)          1.590105   0.138583  11.474   0.0000   1.318488   1.861722
age                -0.013361   0.003780  -3.535   0.0004  -0.020769  -0.005953
type                0.911158   0.071653  12.716   0.0000   0.770720   1.051596
rcs():1            -0.007336   0.028614  -0.256   0.7977  -0.063419   0.048747
rcs():2            -0.159499   0.032838  -4.857   0.0000  -0.223860  -0.095138
rcs():3             0.126317   0.026191   4.823   0.0000   0.074983   0.177651
_cons               3.283403   0.223338  14.701   0.0000   2.845668   3.721138
log_sd(resid.)     -0.578881   0.036576 -15.827   0.0000  -0.650569  -0.507193
fp()               -0.117359   0.072014  -1.630   0.1032  -0.258503   0.023785
_cons               1.377721   0.550711   2.502   0.0124   0.298348   2.457094
log_sd(M2)          1.125207   0.182159   6.177   0.0000   0.768182   1.482232
log_sd(M1)          0.297373   0.240480   1.237   0.2162  -0.173959   0.768704
atanh_corr(M1,M2)  -1.524655   0.257312  -5.925   0.0000  -2.028978  -1.020332

Integration method: Non-adaptive Gauss-Hermite quadrature
Integration points: 9
\end{CodeOutput}
\end{CodeChunk}

The parameters are reported in the results table in the order the
submodels were specified. This model is complex, and it would not be
possible to fit it in the other joint modelling software discussed. This
complexity of the model means it is computationally intensive to
estimate, taking approximately 29 minutes on a 2-core laptop with 8 Gb
of RAM.

\hypertarget{discussion}{%
\section{Discussion}\label{discussion}}

The example above illustrates the flexibility of \pkg{merlin} and the
wide rage of models it is able to fit, the cost of this flexibility is
computational time, more complex models with multiple random effects can
be slow. For particular models it may be possible to calculate the
likelihood more efficiently, which cannot be done here due to the
generalised way \pkg{merlin} has been built. By default the likelihood
is estimated using Gauss-Hermite quadrature to integrate out the
random-effects. For every random effect the likelihood has to be
estimated at each of the quadrature points, meaning for \(r\) random
effects, and \(n\) quadrature points the likelihood has to be estimated
\(n^r\) times. Alternatively the option to use Monte-Carlo integration
is available, which is more efficient, especially for large numbers of
random effects, improving computation time.

A further cost of the flexibility of \pkg{merlin} is the relatively
complex syntax which is necessary to enable the fitting of all possible
models. Whilst the syntax allows for a wide range of models to be
fitted, it could lead to confusion and mistakes in model specification
being made. To address this we have written inbuilt wrapper functions,
such as \code{mlsurv} shown in Section 3.4, and \code{mlrcs}, to fit
specific types of models, which will use the underlying \pkg{merlin}
package, but will have simplified syntax to make them more
user-friendly.

The \pkg{merlin} package is constantly evolving, with many further
updates to \pkg{merlin} planned in line with the \proglang{Stata}
package. Planned updates include implementation of fully adaptive
Gauss-Hermite quadrature, allowing for left truncated survival data, and
empirical Bayes predictions of the random effects.

Overall the flexibility of \pkg{merlin} will allow it to be used to fit
a wide range of new models which cannot be implemented in other software
packages. Its modular nature will allow for the easy addition of further
components and families of outcomes, and the ability to incorporate
user-defined functions means that there are many directions \pkg{merlin}
can be taken in which have not yet been considered.

\hypertarget{acknowledgements}{%
\section{Acknowledgements}\label{acknowledgements}}

This work was supported by the Medical Research Council (MR/P015433/1).

\renewcommand\refname{References}
\bibliography{software.bib}

\begin{thebibliography}{14}
\newcommand{\enquote}[1]{``#1''}
\providecommand{\natexlab}[1]{#1}
\providecommand{\url}[1]{\texttt{#1}}
\providecommand{\urlprefix}{URL }
\expandafter\ifx\csname urlstyle\endcsname\relax
  \providecommand{\doi}[1]{doi:\discretionary{}{}{}#1}\else
  \providecommand{\doi}{doi:\discretionary{}{}{}\begingroup
  \urlstyle{rm}\Url}\fi
\providecommand{\eprint}[2][]{\url{#2}}

\bibitem[{Bates \emph{et~al.}(2015)Bates, M{\"a}chler, Bolker, and
  Walker}]{lme4}
Bates D, M{\"a}chler M, Bolker B, Walker S (2015).
\newblock \enquote{Fitting Linear Mixed-Effects Models Using {lme4}.}
\newblock \emph{Journal of Statistical Software}, \textbf{67}(1), 1--48.
\newblock \doi{10.18637/jss.v067.i01}.

\bibitem[{{Crowther}(2017)}]{Crowther2017}
{Crowther} MJ (2017).
\newblock \enquote{{Extended multivariate generalised linear and non-linear
  mixed effects models}.}
\newblock \emph{arXiv e-prints}, arXiv:1710.02223.
\newblock \eprint{1710.02223}.

\bibitem[{{Crowther}(2018)}]{merlinStata}
{Crowther} MJ (2018).
\newblock \enquote{{merlin - a unified modelling framework for data analysis
  and methods development in Stata}.}
\newblock \emph{arXiv e-prints}, arXiv:1806.01615.
\newblock \eprint{1806.01615}.

\bibitem[{Gould \emph{et~al.}(2015)Gould, Boye, Crowther, Ibrahim, Quartey,
  Micallef, and Bois}]{Gould2015a}
Gould AL, Boye ME, Crowther MJ, Ibrahim JG, Quartey G, Micallef S, Bois FY
  (2015).
\newblock \enquote{Joint modeling of survival and longitudinal non-survival
  data: current methods and issues. Report of the DIA Bayesian joint modeling
  working group.}
\newblock \emph{Statistics in medicine}, \textbf{34}(14), 2181--2195.

\bibitem[{Hickey \emph{et~al.}(2018)Hickey, Philipson, Jorgensen, and
  Kolamunnage-Dona}]{joineRML}
Hickey GL, Philipson P, Jorgensen A, Kolamunnage-Dona R (2018).
\newblock \enquote{joineRML: a joint model and software package for
  time-to-event and multivariate longitudinal outcomes.}
\newblock \emph{BMC Med Res Methodol}, \textbf{18}(1), 50.
\newblock \doi{10.1186/s12874-018-0502-1}.
\newblock \urlprefix\url{http://www.ncbi.nlm.nih.gov/pubmed/29879902}.

\bibitem[{Lim \emph{et~al.}(2008)Lim, Ali, Theodorou, Sousa, Ashrafian,
  Chamageorgakis, Duncan, Henein, Diggle, and Pepper}]{Lim2008}
Lim E, Ali A, Theodorou P, Sousa I, Ashrafian H, Chamageorgakis T, Duncan A,
  Henein M, Diggle P, Pepper J (2008).
\newblock \enquote{Longitudinal study of the profile and predictors of left
  ventricular mass regression after stentless aortic valve replacement.}
\newblock \emph{Ann Thorac Surg}, \textbf{85}(6), 2026--9.
\newblock \doi{10.1016/j.athoracsur.2008.02.023}.
\newblock \urlprefix\url{http://www.ncbi.nlm.nih.gov/pubmed/18498814}.

\bibitem[{Philipson \emph{et~al.}(2018)Philipson, Sousa, Diggle, Williamson,
  Kolamunnage-Dona, Henderson, and Hickey}]{joineR}
Philipson P, Sousa I, Diggle PJ, Williamson P, Kolamunnage-Dona R, Henderson R,
  Hickey GL (2018).
\newblock \emph{{joineR}: Joint Modelling of Repeated Measurements and
  Time-to-Event Data}.
\newblock R package version 1.2.4,
  \urlprefix\url{https://github.com/graemeleehickey/joineR/}.

\bibitem[{Pinheiro \emph{et~al.}(2019)Pinheiro, Bates, DebRoy, Sarkar, and {R
  Core Team}}]{nlme}
Pinheiro J, Bates D, DebRoy S, Sarkar D, {R Core Team} (2019).
\newblock \emph{{nlme}: Linear and Nonlinear Mixed Effects Models}.
\newblock R package version 3.1-141,
  \urlprefix\url{https://CRAN.R-project.org/package=nlme}.

\bibitem[{Rizopoulos(2010)}]{JM}
Rizopoulos D (2010).
\newblock \enquote{JM: An R Package for the Joint Modelling of Longitudinal and
  Time-to-Event Data.}
\newblock \emph{Journal of Statistical Software}, \textbf{35}(9), 1--33.

\bibitem[{Rizopoulos(2016)}]{JMbayes}
Rizopoulos D (2016).
\newblock \enquote{The R Package JMbayes for Fitting Joint Models for
  Longitudinal and Time-to-Event Data Using MCMC.}
\newblock \emph{Journal of Statistical Software}, \textbf{72}(7), 1--46.
\newblock ISSN 1548-7660.
\newblock \doi{10.18637/jss.v072.i07}.
\newblock Ed7td Times Cited:12 Cited References Count:52, \urlprefix\url{<Go to
  ISI>://WOS:000389072600001}.

\bibitem[{Rondeau \emph{et~al.}(2012)Rondeau, Mazroui, and
  Gonzalez}]{frailtypack}
Rondeau V, Mazroui Y, Gonzalez JR (2012).
\newblock \enquote{frailtypack: An R Package for the Analysis of Correlated
  Survival Data with Frailty Models Using Penalized Likelihood Estimation or
  Parametrical Estimation.}
\newblock \emph{Journal of Statistical Software}, \textbf{47}(4), 1--28.
\newblock ISSN 1548-7660.
\newblock 939iw Times Cited:56 Cited References Count:25, \urlprefix\url{<Go to
  ISI>://WOS:000303804000001}.

\bibitem[{Royston(2001)}]{Royston2001}
Royston P (2001).
\newblock \enquote{Flexible Parametric Alternatives to the Cox Model, and
  more.}
\newblock \emph{The Stata Journal}, \textbf{1}(1), 1--28.
\newblock \doi{10.1177/1536867X0100100101}.
\newblock \eprint{https://doi.org/10.1177/1536867X0100100101},
  \urlprefix\url{https://doi.org/10.1177/1536867X0100100101}.

\bibitem[{Therneau(2019)}]{coxme}
Therneau TM (2019).
\newblock \emph{coxme: Mixed Effects Cox Models}.
\newblock R package version 2.2-14,
  \urlprefix\url{https://CRAN.R-project.org/package=coxme}.

\bibitem[{Umlauf \emph{et~al.}(2017)Umlauf, Klein, and Zeileis}]{BAMLSS}
Umlauf N, Klein N, Zeileis A (2017).
\newblock \enquote{{BAMLSS}: {B}ayesian Additive Models for Location, Scale and
  Shape (and Beyond).}
\newblock \emph{Journal of Computational and Graphical Statistics}.

\end{thebibliography}

\end{document}